\documentclass{article}
\input{preamble.sty}
\title{Six-loop renormalization group analysis of the $\ph^4 + \ph^6$ model}

\author[1,2]{L. Ts. Adzhemyan}
\author[1,2]{M. V. Kompaniets}
\author[1,2]{A. V. Trenogin}

\affil[1]{\it Saint Petersburg State University, 7/9 Universitetskaya nab., St. Petersburg 199034, Russian Federation}
\affil[2]{\it Bogolyubov Laboratory of Theoretical Physics, Joint Institute for Nuclear Research, Dubna 141980, Russian Federation}

\date{}

\begin{document}
\maketitle

\begin{abstract}

    We investigate the $\lambda\varphi^4+g\varphi^6$ model using the renormalization group method and the $\varepsilon$ expansion. This model is used in a situation  where the coefficients $\lambda$, $g$ and the coefficient $\tau$ of the term $\tau \varphi^2$ depend on two parameters $T$ and $P$, and there is a point ($T_c,P_c$) at which $\tau$ and $\lambda$ are zero. This point is named the tricritical point. The description of a system depends on a trajectory that leads to the tricritical point on the plane ($T,P$). In the trajectories, when $\lambda$ goes to zero fast enough, the description is defined by the $\varphi^6$ interaction and then the $\varphi^4$ term can be considered as a composite operator. In this case, the logarithmic dimension is $d=3$, and the $\varepsilon$ expansion is carried out in the dimension $d=3-2\varepsilon$.
    The main exponents of the \textit{tricritical} model have been calculated in the third order of the $\varepsilon$ expansion. 
    Taking into account the $\varphi^4$ interaction, we were able to calculate the parameter that determines the required decrease rate in $\lambda$ to implement the tricritical behavior. The tricritical dimensions of the composite operators $\varphi^k$ for $k=1, 2, 4, 6$ have been computed.
    The resulting values are compared to those known from a conformal field theory and non-perturbative renormalization group.

\end{abstract}

\section{Introduction}

%    \todo{цвет не убирать, чтобы я знал, где править в русской версии}
    The study of the tricritical phase transition within the Ginzburg-Landau model is based on the addition of the contribution $\sim\ph^6$ to the action of the $\ph^4$ model:
    \begin{equation}\label{unrenormalized_action}
        S(\ph) = -\frac{1}{2} (\partial \ph)^2 - \frac{1}{2} \tau_0 \ph^2 - \frac{1}{4!} \lambda_{0} \ph^4 - \frac{1}{6!} g_{0} \ph^6.
    \end{equation}
    This term must be considered when there is a line of critical points, $f(T, P)=0$, on the plane defined by the parameters $T$ and $P$ in a system. On this line, the coefficient $\tau_0$ turns to zero, and at its endpoint $(T_c,P_c)$, the equality $\lambda_0(T_c,P_c)=0$ is also fulfilled. In the vicinity of this \textit{tricritical} point, the term $\sim \ph^6$ can play a significant role. 
    Experimental conditions define a path in the $(T, P)$ plane as we approach the tricritical point. For example, $\lambda_0 = \lambda_0(\tau_0)$, if we choose $\tau_0$ as a trajectory parameter. The trajectory can be chosen in such a way that $\tau_0$ and $\lambda_0$ will become values of different order of smallness ($\lambda_0=h_0 \cdot \tau_0^b$). On such trajectories, the relative role of the interactions $\ph^4$ and $\ph^6$ changes depending on the value of $b$. It may be shown \cite{vasiliev} that there is the value $b=b_0$ which determines the type of behavior in the asymptotics $\tau_0 \to 0$. Tricritical behavior is realized for $b>b_0$, since the $\ph^4$ interaction is irrelevant and must be neglected. For $b \leq b_0$, two additional scenarios arise. If the $\ph^6$ interaction becomes irrelevant, modified critical behavior takes place. In this situation, the exponents are simply expressed in terms of $b$ and the exponents of the $\ph^4$ model~\cite{vasiliev,a:Borin:1992}. Otherwise, when both interactions are relevant and must be taken into account, combined tricritical behavior occurs.

    We are interested in the $\ph^4+\ph^6$ model \eqref{unrenormalized_action} where the main interaction is $\ph^6$ and $\ph^4$ is a composite operator. The model investigation is carried out with the help of the renormalization group (RG) method in $d=3-2\ep$ where $\ep>0$ is a logarithmic dimension deviation. As a result of the calculation, we have tricritical exponents in the form of series in $\ep$ \cite{vasiliev}.  Independent calculations of the $\ep$ series were performed by different authors \cite{vasiliev, STEPHEN1973, LewisAdams, Pisarski1982,Hager99, Hager02}, where in the studies \cite{vasiliev, LewisAdams, Hager99, Hager02} the record six-loop (third order of $\ep$ expansion) results were published. These works demonstrate two features of the model that significantly complicate its renormalization compared to the $\ph^3$ and $\ph^4$ models. The first feature is that the $\ph^4+\ph^6$ model is characterized by a faster growth in the number of loops relative to the order of the $\ep$ expansion. To calculate the $l$-th order correction, it is necessary to compute all diagrams up to and including $2l$ loops. The second feature is that high-order asymptotic terms grow faster \cite{Hager99, Hager02} ($\sim(2k)!$ instead of $k!$) than in standard theories such as $\ph^3$ and $\ph^4$. It should also be noted that the most advanced methods for calculating multi-loop diagrams, the hyperlogarithm method \cite{panzerPhD15, a:Panzer2017} and the graphical function method \cite{a:Borinsky:2021jdb, a:Borinsky:2022lds, a:Schnetz:2022nsc}, cannot be used for models with odd logarithmic dimension.

    The feasibility of the tricritical behavior is determined by two values. The first value is the critical exponent $\omega$, which determines the stability of a fixed point. The second is the parameter $b_0$, which defines the type of behavior  (tricritical, modified critical, or combined tricritical behavior) depending on the chosen trajectory. Our goals are to check the previous results, compute the exponent $\omega$ in the third order of $\ep$ expansion and calculate tricritical dimensions of composite operators whose values are known from a conformal field theory and non-perturbative RG \cite{BELAVIN1984333, a:bootstrap2021, Henriksson:2025kws, a:frg_Codello_2012, a:frg_Codello_2015}.

    The paper is organized as follows. Section~\ref{sec:model_and_renorm} contains the model renormalization and the main renormalization group relations. We present our results in Section~\ref{sec:results}. Section~\ref{subsec:previous_results} is a short summary of previous results that were obtained with the help of the renormalization group method and the $\ep$ expansion to date. In Section~\ref{subsec:our_results} we present our results of a six-loop calculation (third order in $\ep$ expansion) and compare them with the results of previous studies. The calculated values of the tricritical exponents and the tricritical dimensions of composite operators in various space dimensions are presented in Section~\ref{subsec:resummation}. In this Subsection, we also compare the computed values of tricritical dimensions of composite operators with those known from a conformal field theory and non-perturbative RG. Final remarks are given in Conclusion~\ref{sec:conclusion}.

\section{Model and renormalization}\label{sec:model_and_renorm}

    We investigate a generalization of the $\ph^6$ \cite{vasiliev} scalar model with the additional $\ph^4$ interaction. The unrenormalized action of the $\ph^4+\ph^6$ model can be written as \eqref{unrenormalized_action}
    with two parameters $\tau_0$ and $\lambda_0$ and a coupling constant $g_0$.
    The renormalized action is
    \begin{equation}\label{renormalized_action}
        S_R(\ph) = -\frac{Z_1}{2}(\partial \ph)^2 - \frac{Z_2}{2}\tau \ph^2 - \frac{Z_4}{4!} {\lambda} \mu^{2\ep} \ph^4 - \frac{Z_3}{6!} g \mu^{4\ep} \ph^6
    \end{equation}
    with the renormalization mass $\mu$. Each insertion of the $\ph^4$ interaction decreases the ultraviolet (UV) divergence of the diagrams. As a consequence, diagrams that have more than two insertions of the $\ph^4$ interaction are not UV-divergent and do not contribute to the renormalization constants. Thus, it is enough to account for the diagrams without the $\ph^4$ interaction to find the renormalization constants of the $\ph^6$ model, and with one or two insertions, to obtain $Z_4$ and $Z_2$, respectively, during the renormalization procedure of the $\ph^4+\ph^6$ model. Because of this, the renormalization constants $Z_1, ~Z_3$ and $Z_4$ depend only on the renormalized coupling constant $g$, but $Z_2$ has the form $Z_2(g,\lambda^2/\tau) = Z'_2(g) + \delta Z_2(g) \lambda^2/\tau$, in which $Z_2'$ is a renormalization constant of the $\ph^6$ model. The transition from \eqref{unrenormalized_action} to \eqref{renormalized_action} is carried out using multiplicative renormalization of the field, parameters, and coupling constant
    \begin{equation}\label{param_renorm}
        \hat\ph = Z_{\ph} \hat\ph_R, \quad \tau_0 = Z_{\tau} \tau = \left( Z_\tau'+ \delta{Z_\tau} \frac{v^2}{\tau} \right) \tau, \quad \lambda_{0} = c Z_{\lambda} v \mu^{2\ep}, \quad g_{0} = c^2 Z_{g} u \mu^{4\ep},
    \end{equation}
    where we introduce a normalized coupling constant $u=g/c^2$ and a parameter $v=\lambda/c$ with $c$ from \eqref{c}, $Z'_\tau$ is a renormalization constant of the $\ph^6$ model.

    We use the dimensional regularization \cite{regularization} (in $d=3-2\ep$ Euclidean dimensions) and the specific realization of the minimal subtraction scheme, the G-scheme \cite{GFunctions}. In that case, counterterms are presented as poles by $\ep$, and the normalization of the coupling constant $u$ and the parameter $v$ is chosen so that the sunset diagram is equal to $1/\ep$. For this, $c$ is selected as
    \begin{align}\label{c}
        c &= \sqrt{(4\pi)^{(3-2\ep)} \cdot \frac{\Gamma(\frac{3}{2}-3\ep)}{\Gamma^3(\frac{1}{2}-\ep) \Gamma(2\ep)} \cdot \frac{1}{\ep}} = 8\pi + \mathcal{O}(\ep).
    \end{align}
    An additional advantage of using the G-scheme is a simpler form of divergencies of diagrams.
    
    The renormalization constants \eqref{renormalized_action} and \eqref{param_renorm} are connected by relations
    \begin{equation}
        Z_1 = Z_{\ph}^2, \quad Z_2 = Z_{\tau} Z_{\ph}^2,   \quad Z_4 = Z_{\lambda} Z_{\ph}^4, \quad Z_3 = Z_{g} Z_{\ph}^6. \label{eq:rc_relations}
    \end{equation}
    With the help of the renormalization constants, one can calculate four independent RG functions~\cite{vasiliev}
    \begin{equation}
        \gamma_i(u) = -4 \, u \, \partial_u Z_i^{(1)} |_{i\in\{1,3,4\}}, \qquad \gamma_2 \left( u,\frac{v^2}{\tau} \right) = -4 \, \left( u \, \partial_u Z_2'^{(1)} + \frac{v^2}{\tau} \left( u \, \partial_u + 1 \right) \delta Z_2^{(1)} \right), \label{eq:gamma1234}
    \end{equation}
    where $Z_i^{(1)}$ is a coefficient of the first pole in $Z_i$. Both equations are obtained from the condition that RG functions must be UV-convergent \cite{vasiliev}. Using $\gamma_i|_{i \in \{1,2,3,4\}}$ and the relations between the renormalization constants \eqref{eq:rc_relations} allows us to introduce and calculate RG functions of the field, parameters, and coupling constant:
    \begin{equation}
        \gamma_\ph (u)= \frac{\gamma_1}{2};
        \quad \gamma_\lambda(u) = \gamma_4-2\gamma_1;
        \quad \gamma_u (u)= \gamma_3-3\gamma_1;
        \quad \gamma_\tau \left( u,\frac{v^2}{\tau} \right) \equiv \gamma_\tau'(u) + \delta \gamma_\tau(u) \frac{v^2}{\tau} = \gamma_2-\gamma_1. \label{eq:gamma_phi_lambda_tau_u}
    \end{equation}
    Now we can present $\beta$ function \cite{vasiliev} through introduced $\gamma_u$ using the next relation
    \begin{equation}
        \beta(u) = -u (4\ep + \gamma_u).\label{eq:beta}
    \end{equation}
    
    The fixed points $\{u_*(\ep)\}$ are determined by the roots of the equation $\beta(u_*)=0$. We are interested in the infrared (IR) asymptote which is defined by a non-trivial fixed point. The IR stability of this point corresponds to the condition
    \begin{equation}
        \omega \equiv \partial_u\beta |_{u=u_*} > 0.\label{eq:omega}
    \end{equation}
    Values of the RG functions $\gamma_\ph$ and $\gamma_\tau'$ in the fixed point $u_*$
    \begin{equation}
        \gamma_{\varphi}^*(\ep) \equiv \gamma_{\varphi}(u_*) = \frac{\gamma_1(u_*)}{2} \qquad \text{and} \qquad \gamma_\tau'^*(\ep) \equiv \gamma_\tau'(u_*) = \gamma_2(u_*)|_{v^2=0} - \gamma_1(u_*) \label{eq:gamma_phi_and_tau}
    \end{equation}
    define two independent tricritical exponents of the $\ph^6$ model, which can be selected as the Fisher exponent
    \begin{equation}
        \eta = 2\gamma_\ph^* \label{eq:eta}
    \end{equation}
    and the tricritical exponent of the correlation length
    \begin{equation}
        \nu = (2+\gamma_\tau'^*)^{-1}. \label{eq:nu}
    \end{equation}
    Taking into account the additional $\ph^4$ interaction to the main $\ph^6$ interaction allows us to calculate $\gamma_\lambda^*$. It can be shown \cite{vasiliev, Henriksson:2025kws} that the tricritical dimensions of the field and composite operators ${\ph^{2i}}|_{i\in\{1,2,3\}}$ are connected with the investigated values ($\gamma_\ph^*, ~\gamma_\tau'^*, ~\gamma_\lambda^*$ and $\omega$) through the following relations:
    \begin{align}
        \Delta_{\ph^k} &= k \left( \frac{1}{2} - \ep \right) + \gamma_{\ph^k}^*; \qquad
        \gamma_{\ph^2}^* = - \gamma_{\tau}'^*; \qquad \gamma_{\ph^4}^* = -\gamma_{\lambda}^*; \qquad \gamma_{\ph^6}^* = 4\ep + \omega, \label{eq:composite_dimensions}
    \end{align}
    exact values of which are calculated using the conformal field theory in $d=2$ \cite{BELAVIN1984333, Henriksson:2025kws}:
    \begin{equation}
        \Delta_{\ph}|_{d=2} = \frac{3}{40}; \qquad \Delta_{\ph^2}|_{d=2} = \frac{1}{5}; \qquad \Delta_{\ph^4}|_{d=2} = \frac{6}{5}; \qquad \Delta_{\ph^6}|_{d=2} = 3.
        \label{eq:exact_dimensions}
    \end{equation}
    Recently, a new semiclassical approach has been proposed to compute $\Delta_{\ph^k}$ in the limit of large $k$ \cite{a:antipin_1, a:antipin_2}.
    
    The tricritical behavior is implemented when the parameter $v$ goes to zero with respect to the parameter $\tau$ at a fast enough rate. Assuming $v = h \cdot \tau^b$, one can determine the parameter $b_0$ that separates two regions of distinct behavior. Tricritical behavior is realized in the case of $b > b_0$. The parameter $b_0$ is expressed in terms of $\nu$ and $\gamma_\lambda^*$ \cite{vasiliev} with the help of the following relation
    \begin{equation}
        b_0 = \nu \cdot (1 + 2 \ep + \gamma_{\lambda}^*). \label{eq:b0}
    \end{equation}
    For the trajectories with $b < b_0$, the behavior depends on the sign of the parameter $a$ introduced in the RG analysis of the $\ph^4+\ph^6$ model in \cite{vasiliev}\footnote{Additional work is required to assess the universality of $a$.}
    \begin{equation}
        a = \frac{\delta \gamma_\tau^*}{\gamma_\tau'^*-4\ep-2\gamma_\lambda^*} = \frac{\nu}{1-2b_0} \cdot \delta \gamma_\tau^*.
        \label{a}
    \end{equation}
    Сombined tricritical behavior occurs for $a<0$, and modified critical behavior otherwise. The trajectories with value $b=b_0$ are unstable.
    
\section{Results}\label{sec:results}

    \subsection{Previous results}\label{subsec:previous_results}
        \begin{table}[h!]
            \centering
            \captionsetup{justification=centering}
            \begin{tabular}{lllll|l}
                Sources & $\eta$ & $\nu$ & $\omega$ &  $b_0$ & $\delta Z_2(u) \frac{v^2}{\tau}$\\
                \hline
                 \cite{STEPHEN1973} (1973) & 2 & 2 & - & - & -\\
                \cite{LewisAdams} (1978) & 3  & 2 & - & 2 & -\\
                \cite{vasiliev} (rus1998, eng2004)  &3 & 3 & 2$^{(1)}$ & 1 & 1 \\
                \cite{Hager99} (1999) & 3 & 3 & 2$^{(1)}$ & - & -\\
                 \cite{Hager02} (2002) & 3 & 3 & 3$^{(1,2)}$ & 3$^{(2)}$ & -\\
                 \hline
                current work & 3 & 3 & 3 & 3 & 3\\
            \end{tabular}
            \caption{
            Number of terms of the $\ep$ expansion of the $\ph^4+\ph^6$ model calculated in the papers for each quantity (or its equivalents).  The last column shows an order in $\left(u,\frac{v^2}{\tau}\right)$ calculated in the papers.
            First three columns ($\eta, ~\nu, ~\omega$) correspond to the $\ph^6$ model, the last two -- to the $\ph^4+\ph^6$ model with the main $\ph^6$ and the additional $\ph^4$ interactions. In the studies \cite{vasiliev, Hager99, Hager02}, the exponent $\omega$ is not calculated. Nevertheless, expressions for the $\beta$ function are presented there and can be used to compute $\omega$ in the corresponding order (marked by $^{(1)}$). Results that contain error marked by $^{(2)}$.
            }
            \label{tab:indices}
        \end{table}

        The investigation of the $\ph^4+\ph^6$ model began as soon as Wilson realized the possibility of using the renormalization group method to study critical behavior. We would like to point out that in all the works mentioned below, with the exception of the monograph \cite{vasiliev}, the $O(n)$-symmetric generalization of the $\ph^4+\ph^6$ model is investigated. The first terms of the series in $\ep$ ($\sim \ep^2$) of the tricritical exponents are presented in \cite{STEPHEN1973} (1973). A correction $\sim \ep^3$ in $\eta$ is submitted in \cite{LewisAdams} (1978). In the same paper, contributions of the order of $\ep$ and $\ep^2$ to the parameter $b_0$ are given\footnote{$b_0$ is denoted as $\alpha_0$ in \cite{vasiliev}, as $\phi_t$ in \cite{LewisAdams} and as $\ph$ in \cite{Hager02}}. In the monograph \cite{vasiliev}  a term $\sim\ep^3$ of $\gamma_\tau'^*$ is presented. The term can be used to calculate $\nu$ in the third order. Additionally, there is an expression for the $\beta$ function, which allows us to compute $\omega$ up to $\ep^2$. Also, in \cite{vasiliev} $\delta Z_2$ is calculated in leading order and its contribution to the renormalization group equation is discussed. The article \cite{Hager99} contains the $O(n)$-symmetric generalization of the term $\sim \ep^3$ in $\nu$ and the expression for the $\beta$ function. In addition, there is $O(n)$-symmetric expression for $\eta$ up to $\ep^3$, which independently confirms the result of the work \cite{LewisAdams}. In the work \cite{Hager02}, all the available results to date can be found, including the correction $\sim \ep^3$ into $b_0$. The result of the $\beta$ function allows one to calculate $\omega$ up to the third order. This confirms the first two terms of the series from \cite{vasiliev}. For clarity, we present the same results in Table \ref{tab:indices}.

        When reviewing the article \cite{Hager02}, we had several questions related to the inaccuracies found therein.
        In the next section, we provide the results of our calculation and compare them with \cite{Hager02}.

    \subsection{Our results}\label{subsec:our_results}
    
        \subsubsection{General results}

            We calculated all the necessary diagrams up to six loops, both analytically and numerically. For analytic calculation of the complicated diagrams (which include a
            t-bubble\footnote{T1 topology from~\cite{Larin:1991fz}}~\cite{a:t-bubble} diagram with non-integer indices), we use results  for t-bubble diagrams provided by A. Pikelner\footnote{Unpublished results of A. Pikelner.}. For all diagrams, the numerical and analytical results match perfectly. We use analytical expressions to obtain the final results in analytical form.
        
            Using the diagrams calculated up to six-loop, we can find the renormalization constants:
            \begin{align}
                Z_1 &= 1 - \frac{1}{360} \frac{u^2}{\ep} + \frac{1}{972} \Bigl[ \frac{4}{\ep} - \frac{3}{\ep^2} \Bigr] u^{3} + \mathcal{O}(u^4); \label{expr:Z1} \\
                %%%
                Z_2' &= 1 + \frac{1}{24} \frac{u^2}{\ep} - \frac{1}{216} \Bigl[ (88+3\pi^{2}) \frac{1}{\ep} - \frac{10}{\ep^{2}} \Bigr] u^{3} + \mathcal{O}(u^4); \label{expr:Z2} \\
                %%%
                \delta Z_2 &= \frac{1}{6\ep} - \frac{1}{288} \Bigl[ (32+3\pi^{2}) \frac{3}{\ep} - \frac{32}{\ep^{2}} \Bigr] u + \frac{1}{12960} \Bigl[ \Bigl\{ 8 \Bigl( 1018 - 189\zeta(3) + 864\beta_D(4) \Bigr) \nonumber \\
                &+ 36\pi^{2} \Bigl( 17 + \ln(2) + 8G \Bigr) + 81\pi^{4} \Bigr\} \frac{5}{\ep} - (3746+315\pi^{2}) \frac{3}{\ep^{2}} + \frac{1440}{\ep^{3}} \Bigr] u^{2} + \mathcal{O}(u^3); \label{expr:delta_Z2} \\
                %%%
                Z_3 &= 1 + \frac{5}{3} \frac{u}{\ep} - \frac{5}{144} \Bigl[ (10+\pi^{2}) \frac{27}{\ep} - \frac{80}{\ep^{2}} \Bigr] u^{2} + \frac{1}{1296} \Bigl[ \Bigl\{ 8 \Bigl( 4372 - 2835\zeta(3) + 7776\beta_D(4) \Bigr) \nonumber \\
                &+ 6 \pi^{2} \Bigl( \textcolor{orange}{619 + 270\ln(2)} + 432G \Bigr) + 405\pi^{4} \Bigr\} \frac{5}{\ep} - (15746+1575\pi^{2}) \frac{3}{\ep^{2}} + \frac{6000}{\ep^{3}} \Bigr] u^{3} + \mathcal{O}(u^4); \label{expr:Z3} \\
                %%%
                Z_4 &= 1 + \frac{2}{3} \frac{u}{\ep} - \frac{1}{144} \Bigl[ (116+9\pi^{2}) \frac{3}{\ep} - \frac{112}{\ep^{2}} \Bigr] u^{2} + \frac{1}{3240} \Bigl[ \Bigl\{ 16 \Bigl( 1334 - 567\zeta(3) + 1296\beta_D(4) \Bigr) \nonumber \\
                &+ \pi^{2} \Bigl(\textcolor{orange}{1303 + 864\ln(2)} + 864G \Bigr) + 162\pi^{4} \Bigr\} \frac{5}{\ep} - (4568+405\pi^{2}) \frac{6}{\ep^{2}} + \frac{3360}{\ep^{3}} \Bigr] u^{3} + \mathcal{O}(u^4), \label{expr:Z4}
            \end{align}
            where $\beta_D$ and $\zeta$ -- Dirichlet beta function and Riemann zeta function, $G \equiv \beta_D(2)$ and $\gamma_E$ -- Catalan's and Euler's constants. In these expressions and in those that follow, some terms are highlighted in a different color. These terms are different comparing to \cite{Hager02} and will be discussed in Sec.~\ref{subsubsec:comparison}.
        
            Using the calculated renormalization constants and the RG functions \eqref{eq:gamma1234}, \eqref{eq:gamma_phi_lambda_tau_u}, \eqref{eq:beta} we obtain:
            \begin{align}
                \beta(u) &= -4 \ep u + \frac{20}{3}u^{2} - \frac{(2248+225\pi^{2})}{30} u^{3} + \frac{1}{36} \Bigl[ 8 \Bigl( 7286 - 4725\zeta(3) + 12960\beta_D(4) \Bigr) \nonumber \\
                &+ 10\pi^{2} \Bigl( \textcolor{orange}{619 + 270\ln(2)} + 432G \Bigr) + 675\pi^{4} \Bigr] u^{4} + \mathcal{O}(u^5); \label{expr:beta} \\
                %%%
                \gamma_\varphi(u) &= \phantom{-} \frac{1}{90} u^2 - \frac{2}{81} u^3 + \mathcal{O}(u^4); \label{expr:rg_gamma_phi} \\
                %%%
                \gamma_\tau'(u) &= -\frac{16}{45} u^{2} + \frac{(800+27\pi^{2})}{162} u^{3} + \mathcal{O}(u^4); \label{expr:rg_gamma_tau} \\
                %%%
                \gamma_\lambda(u) &= -\frac{8}{3} u + \frac{(1736+135\pi^{2})}{90} u^{2} - \frac{1}{162} \Bigl[ 16 \Bigl( 4001-1701\zeta(3)+3888\beta_D(4) \Bigr) \nonumber \\
                &+ 3\pi^{2} \Bigl( \textcolor{orange}{1303+864\ln(2)} + 864G \Bigr) + 486\pi^{4} \Bigr] u^{3} + \mathcal{O}(u^4); \label{expr:rg_gamma_lambda} \\
                %%%
                \delta \gamma_\tau(u) &= -\frac{2}{3} + \frac{(32+3\pi^{2})}{12} u - \frac{1}{216} \Bigl[ 8 \Bigl( 1018-189\zeta(3)+864\beta_D(4) \Bigr) \nonumber \\
                &+ 36\pi^{2} \Bigl( 17+\ln(2)+8G \Bigr) + 81\pi^{4} \Bigr] u^{2} + \mathcal{O}(u^3) .\label{expr:rg_delta_gamma_tau} 
            \end{align}
        
            From \eqref{expr:beta} a fixed point $u_*$ ($\beta(u_*) = 0$) is obtained:
            {\begin{align}
                u_* &= \frac{3}{10} (2\ep) + \frac{9}{20000}(2248+225\pi^{2}) (2\ep)^{2} + \frac{9}{20000000} \Bigl\{ 16 \Bigl( 36782+590625\zeta(3)-1620000\beta_D(4) \Bigr) \nonumber \\
                &+ 100\pi^{2} \Bigl( \textcolor{orange}{14873-6750\ln(2)} - 10800G \Bigr) - 16875\pi^{4} \Bigr\} (2\ep)^{3} + \mathcal{O}(\ep^4). \label{expr:fp}
            \end{align}

            The exponent $\omega$ \eqref{eq:omega}, which is responsible for the IR stability of the fixed point, is expressed as follows:
            \begin{align}
                \omega &= 2 \cdot (2\ep) - \frac{3}{1000}(2248+225\pi^{2}) (2\ep)^{2} + \frac{3}{1000000} \Bigl\{ 32 \Bigl( 436984-590625\zeta(3)+1620000\beta_D(4) \Bigr) \nonumber \\
                &+ 200\pi^{2} \Bigl( \textcolor{orange}{301+6750\ln(2)} + 10800G \Bigr) + 185625\pi^{4} \Bigr\} (2\ep)^{3} + \mathcal{O}(\ep^4). \label{expr:omega}
            \end{align}
            
            Using \eqref{expr:fp} and \eqref{eq:eta}, \eqref{eq:nu}, \eqref{eq:b0}, \eqref{expr:rg_delta_gamma_tau} we obtain the exponents $\eta$ and $\nu$
            \begin{align}
                \eta &= \frac{1}{500} (2\ep)^{2} + \frac{1}{1500000}(18232+2025\pi^{2}) (2\ep)^{3} + \mathcal{O}(\ep^4); \label{expr:eta} \\
                %%%
                \nu &= \frac{1}{2} + \frac{1}{125} (2\ep)^{2} + \frac{1}{3000000}(61856+12825\pi^{2}) (2\ep)^{3} + \mathcal{O}(\ep^4), \label{expr:nu}
            \end{align}
            and the values $b_0$ and $\delta \gamma_\tau^*$
            \begin{align}
                b_0 &= \frac{1}{2} + \frac{1}{10} (2\ep) - \frac{3}{10000}(1576+225\pi^{2}) (2\ep)^{2} \nonumber \\
                &+ \frac{1}{10000000} \Bigl\{ 288 \Bigl( 6536-118125\zeta(3)+360000\beta_D(4) \Bigr) \nonumber \\
                &- 50\pi^{2} \Bigl( \textcolor{orange}{34547-37800\ln(2)} - 86400G \Bigr) + 151875\pi^{4} \Bigr\} (2\ep)^{3} + \mathcal{O}(\ep^4); \label{expr:b0} \\
                %%%
                \delta \gamma_\tau^* &= -\frac{2}{3} + \frac{(32+3\pi^{2})}{40} (2\ep) - \frac{1}{240000} \Bigl\{ 32 \Bigl( 5218-4725\zeta(3)+21600\beta_D(4) \Bigr) \nonumber \\
                &- 72\pi^{2} \Bigl( 893-50\ln(2)-400G \Bigr) + 2025\pi^{4} \Bigr\} (2\ep)^{2} + \mathcal{O}(\ep^3).
            \end{align}
        
        \subsubsection{Numerical form}

            RG functions are calculated as follows:
            \begin{align}
                \beta(u) &= -4 \ep u + 6.666667 u^{2} - 148.955366 u^{3} + 8326.463576 u^{4} + \mathcal{O}(u^5); \label{numer_expr:beta} \\
                \gamma_\varphi(u) &= \phantom{-} 0.011111 u^2 - 0.024691 u^3 + \mathcal{O}(u^4); \label{numer_expr:rg_gamma_phi} \\
                \gamma_\tau'(u) &= -0.355556 u^2 + 6.583206 u^3 + \mathcal{O}(u^4); \label{numer_expr:rg_gamma_tau} \\
                \gamma_\lambda(u) &= -2.666667 u + 34.093295 u^2 - 1357.447718 u^3 + \mathcal{O}(u^4); \label{numer_expr:rg_gamma_lambda} \\
                \delta \gamma_\tau(u) &= -0.666667 + 5.134068 u - 138.621625 u^2 + \mathcal{O}(u^3). \label{numer_expr:rg_delta_gamma_tau}
            \end{align}
            
            The fixed point and the anomalous dimensions:
            \begin{align}
                u_* &= \phantom{-} 0.3 \cdot (2\ep) + 2.010897 \cdot (2\ep)^{2} - 6.764121 \cdot (2\ep)^{3} + \mathcal{O}(\ep^4); \label{numer_expr:fp} \\
                \gamma_\ph^* &= \phantom{-} 0.001 \cdot (2\ep)^2 + 0.012739 \cdot (2\ep)^3 + \mathcal{O}(\ep^4); \label{numer_expr:gamma_phi} \\
                \gamma_\tau'^* &= -0.032 \cdot (2\ep)^2 - 0.251245 \cdot (2\ep)^3 + \mathcal{O}(\ep^4); \label{numer_expr:gamma_tau} \\
                \gamma_\lambda^* &= -0.8 \cdot (2\ep) - 2.293997 \cdot (2\ep)^2 + 22.521439 \cdot (2\ep)^3 + \mathcal{O}(\ep^4); \label{numer_expr:gamma_lambda} \\
                \delta \gamma_\tau^* &= -0.666667 + 1.54022 \cdot (2\ep) - 2.151862 \cdot (2\ep)^2 + \mathcal{O}(\ep^3). \label{numer_expr:delta_gamma_tau}
            \end{align}
            
            The exponents and $b_0$ are expressed as:
            \begin{align}
                \omega &= 2 \cdot (2\ep) - 13.405983 \cdot (2\ep)^2 + 269.908654 \cdot (2\ep)^3 + \mathcal{O}(\ep^4); \label{numer_expr:omega} \\
                \eta &= 0.002 \cdot (2\ep)^2 + 0.025479 \cdot (2\ep)^3 + \mathcal{O}(\ep^4); \label{numer_expr:eta} \\
                \nu &= 0.5 + 0.008 \cdot (2\ep)^2 + 0.062811 \cdot (2\ep)^3 + \mathcal{O}(\ep^4); \label{numer_expr:nu} \\
                b_0 &= 0.5 + 0.1 \cdot (2\ep) - 1.138998 \cdot (2\ep)^2 + 11.325131 \cdot (2\ep)^3 + \mathcal{O}(\ep^4). \label{numer_expr:b0}
            \end{align}
            
            The parameter $a$ will be calculated using
            \begin{equation}
                a = \frac{\nu}{1-2b_0} \cdot \delta \gamma_\tau^*, \label{a2}
            \end{equation}
            where $\nu,~b_0$ and $\delta \gamma_\tau^*$ will be         resummed separately.

        \subsubsection{Comparison with previous results}\label{subsubsec:comparison}

            Our results fully coincide with the results of the works \cite{vasiliev,STEPHEN1973,LewisAdams,  Hager99}, but we have differences with \cite{Hager02}. Unfortunately, we were unable to determine the source of error in \cite{Hager02}, therefore, we calculated our results in two different ways (analytically and numerically). Both analytical and numerical calculations delivered the same results, reassuring us in the correctness of our approach.
            
            We now check our results for the renormalization constants \eqref{expr:Z1}, \eqref{expr:Z2}, \eqref{expr:Z3} and \eqref{expr:Z4} with the same from \cite{Hager02} in the $n=1$ case. The expressions for $Z_1$ and $Z_2'$ match perfectly, while there are some differences in $Z_3$ ($Z_6$ in the notation of \cite{Hager02}) and $Z_4$. These differences manifest themselves in simple poles containing $\pi^2$ and $\pi^2 \ln(2)$:
            \begin{align}
                Z^H_6 - Z_3 &= \frac{1}{24} \frac{\pi^2}{\ep} \Bigl( 275 - 30 \ln(2) \Bigr) u^3; \\
                Z^H_4 - Z_4 &=  \frac{1}{72} \frac{\pi^2}{\ep} \Bigl( 197 - 24 \ln(2) \Bigr) u^3.
            \end{align}
            Unfortunately, there are no results for the individual diagrams in the study \cite{Hager02}, so we could not analyze this discrepancy more deeply. The difference in the renormalization constants $Z_3$ and $Z_4$ leads to different terms $\sim \ep^3$ at the fixed point, $\omega$ and $\gamma_\lambda^*$, as well as to a different contribution $\sim \ep^3$ in $b_0$ \eqref{eq:b0}. The six-loop expressions for the anomalous dimensions $\eta$ \eqref{expr:eta} and $\nu$ \eqref{expr:nu} fully coincide with the results of the studies \cite{vasiliev,LewisAdams,  Hager99, Hager02}. This is partly due to the fact that, to obtain a six-loop contribution to $\eta$ and $\nu$, it is enough to have only a four-loop expression for the fixed point (see \eqref{expr:rg_gamma_phi}, \eqref{expr:rg_gamma_tau} and \eqref{expr:fp}). In addition, the terms different from those in \cite{Hager02} are highlighted in orange in the preceding expressions.
            It looks like these terms are not related to the most complex t-bubble diagram.

            Unlike the work \cite{Hager02}, several works \cite{vasiliev, Hager99, JackJones} contain results of a part of the diagrams. The monograph \cite{vasiliev} and the article \cite{Hager99} contain results for four-loop and six-loop diagrams, which are necessary to calculate $Z_1$ and $Z_2$ in the third order, and $Z_3$ in the second order. Additionally, an article \cite{JackJones}\footnote{See the last arXiv version where some typos were corrected.} contains results  for eight six-loop diagrams. This allows us to compare a result of each of these diagrams individually, but a full set of six-loop diagrams is absent. 
            Our results are in full agreement with the results of the studies~\cite{ vasiliev, Hager99, JackJones}.

            We also note that the article \cite{Hager02} contains a few typos. For example, the fixed point (see Eq. (22) of  \cite{Hager02}) does not nullify the $\beta$ function (see Eq. (18) of \cite{Hager02}) in the investigated order. The same problem was discussed by Henriksson in \cite{Henriksson:2025kws}. However, correcting these typos is meaningless, since the expressions for the renormalization constants (see Eqs. (11) and (12) of \cite{Hager02}) are incorrect, as discussed above.

    \subsection{Resummation}\label{subsec:resummation}
        To obtain values of the anomalous dimensions in a specific $d$, we need to equate $\ep$ with the corresponding value in the expressions \eqref{numer_expr:gamma_phi} -- \eqref{numer_expr:b0}.
        These series are asymptotically divergent, it is necessary to resum them first. Nevertheless, some of them ($\gamma_\ph^*$ and $\gamma_\tau'^*$) have only two non-trivial terms in the series in $\ep$, so the resummation of them is meaningless. While $\omega$ and $\gamma_\lambda^*$ have three terms, so we resummed them using a Pade approximant [2/1]. We repeat the expressions for $\gamma_\ph^*$ \eqref{numer_expr:gamma_phi} and $\gamma_\tau'^*$ \eqref{numer_expr:gamma_tau} and present the Pade approximants for $\omega$ and $\gamma_\lambda^*$ in the third order of $\ep$ expansion for convenience:
        \begin{align}
            \gamma_\ph^* &\simeq \phantom{-} 0.001 \cdot (2\ep)^2 + 0.012739 \cdot (2\ep)^3 ; \label{resum_expr:gamma_phi} \\
            \gamma_\tau'^* &\simeq -0.032 \cdot (2\ep)^2 - 0.251245 \cdot (2\ep)^3; \label{resum_expr:gamma_tau} \\
            \gamma_\lambda^* &\simeq -\frac{0.8 \cdot (2\ep) + 10.1480 \cdot (2\ep)^2}{1 + 9.8176 \cdot (2\ep)}; \label{resum_expr:gamma_lambda} \\
            \omega &\simeq \phantom{-} \frac{2\cdot(2 \ep) + 26.8609 \cdot (2\ep)^2}{1 + 20.1334 \cdot (2\ep)}. \label{resum_expr:omega}
        \end{align}
        In addition, we resummed $\delta \gamma_\tau^*$ \eqref{numer_expr:delta_gamma_tau} using a Pade approximant [1/1]:
        \begin{equation}
            \delta \gamma_\tau^* \simeq - \frac{0.6667 - 0.6088 \cdot (2\ep)}{1 + 1.3971 \cdot (2\ep)}. \label{resum_expr:delta_gamma_tau}
        \end{equation}
        
        We are mostly interested in $d=2$, so we substitute $\ep=1/2$ (due to the low number of terms, it was not possible to accurately estimate errors \cite{a:Borinsky:2021jdb, Adzhemyan2019}):
        \begin{align}
            \gamma_\ph^*|_{d=2} &\simeq \phantom{-} 0.014; \label{resum:gamma_phi} \\
            \gamma_\tau'^*|_{d=2} &\simeq -0.283; \label{resum:gamma_tau} \\
            \gamma_\lambda^*|_{d=2} &\simeq -1.012; \label{resum:gamma_lambda} \\
            \omega|_{d=2} &\simeq \phantom{-} 1.366; \label{resum:omega} \\
            \delta \gamma_\tau^*|_{d=2} &\simeq -0.024. \label{resum:delta_gamma_tau}
        \end{align}
        The value of $\omega$ \eqref{resum:omega} is positive, so the fixed point \eqref{numer_expr:fp} is IR-stable.

        Using the results above together with the relations \eqref{eq:eta} and \eqref{eq:nu}, we obtain the following estimates for the exponents:
        \begin{align}
            \eta|_{d=2} &\simeq 0.028; \label{resum:eta} \\
            \nu|_{d=2} &\simeq 0.582. \label{resum:nu}
        \end{align}
        To compute $b_0$ and, consequently $a$, in $d=2$, we resum $b_0$ in two different ways. First, we resum \eqref{numer_expr:b0} using a Pade approximant [2/1]:
        \begin{align}
            &b_0 \simeq \frac{0.5 + 5.0715 \cdot (2\ep) - 0.1447 \cdot (2\ep)^2}{1 + 9.9431 \cdot (2\ep)}; \\
            &b_0|_{d=2} \simeq 0.496. \label{resum:b01}
        \end{align}
        Substituting this result into the relation \eqref{a2}, together with \eqref{resum:delta_gamma_tau} and \eqref{resum:nu}, we get
        \begin{equation}
            a|_{d=2} \simeq -1.746. \label{resum:a1}
        \end{equation}
        Alternatively, $b_0$ can be calculated using the previously obtained value of $\gamma_\lambda^*$ \eqref{resum:gamma_lambda} and the relation \eqref{eq:b0}:
        \begin{equation}
            b_0|_{d=2} \simeq 0.575. \label{resum:b02}
        \end{equation}
        Using this estimate for $b_0$, we obtain
        \begin{equation}
            a|_{d=2} \simeq 0.093. \label{resum:a2}
        \end{equation}

        Tricritical dimensions of the composite operators $\Delta_{\ph^k}$ for $k=1,2,4,6$ {were} approximately calculated using the relations \eqref{eq:composite_dimensions}, and the results may be compared with the exact values \eqref{eq:exact_dimensions}:
        \begin{align}
            \Delta_\ph|_{d=2} &= \left[\frac{1}{2} - \ep + \gamma_{\ph}^*\right] |_{d=2} \simeq 0.014 \qquad \left(\text{the exact value is } 3/40 \right); \label{resum:composite_operator_phi} \\
            \Delta_{\ph^2}|_{d=2} &= [1 - 2\ep - \gamma_{\tau}'^*] |_{d=2} \hspace{0.07cm} \simeq 0.283 \qquad \left( 1/5 \right); \label{resum:composite_operator_phi2} \\
            \Delta_{\ph^4}|_{d=2} &= [2 - 4\ep - \gamma_{\lambda}^*] |_{d=2} \hspace{0.15cm}\simeq 1.012 \qquad \left( 6/ 5\right); \label{resum:composite_operator_phi4} \\
            \Delta_{\ph^6}|_{d=2} &= [3 - 2\ep + \omega] |_{d=2} \hspace{0.29cm} \simeq 3.366 \qquad \left( 3 \right). \label{resum:composite_operator_phi6}
        \end{align}
        Note that the canonical dimension of the composite operator $\ph^k$ is zero in $d=2$ ($\ep=1/2$) for any value of $k$. Thus, the results \eqref{resum:composite_operator_phi} -- \eqref{resum:composite_operator_phi6} are fully formed by the anomalous dimensions. For clarity, we present the same results in Table \ref{tab:composite_dim}.

        \renewcommand*{\arraystretch}{1.4}
        \begin{table}[h!]
        \centering
        \captionsetup{justification=centering}
            \begin{tabular}{c|c|c}
%                \hline
                $k$ & {this work} & exact \\
                \hline
                1 & 0.014 & 0.075 \\
                2 & 0.283 & 0.2 \\
                4 & 1.012 & 1.2 \\
                6 & 3.366 & 3 \\
%                \hline
            \end{tabular}
            \caption{
            Tricritical dimensions of composite operators $\Delta_{\ph^k}$ \eqref{eq:composite_dimensions} in $d=2$ ($\ep=1/2$). \\
            {The second column contains approximate values that were obtained using \eqref{resum_expr:gamma_phi} -- \eqref{resum_expr:omega}}, while the third one contains the exact values which are given by the conformal field theory \cite{BELAVIN1984333, Henriksson:2025kws}.
            }
            \label{tab:composite_dim}
        \end{table}

        In addition, we know about two studies \cite{a:bootstrap2021, Henriksson:2025kws} in which $\Delta_\ph$ and $\Delta_{\ph^2}$ are investigated in $2 < d < 3$ using conformal bootstrap. In \cite{a:bootstrap2021} one can find a comparison with the results that can be obtained from the $\ep$ expansion for $\Delta_\ph$. For comparison, they use the result of work \cite{LewisAdams}, so the values from \eqref{resum_expr:gamma_phi} are the same. In \cite{Henriksson:2025kws} Henriksson also calculates $\Delta_{\ph^2}$ for $d = 2.5 \text{ and } 2.75$, and compares his results with those obtained using a non-perturbative RG in \cite{a:frg_Codello_2012}. The corresponding values from the $\ep$ expansion can be obtained from \eqref{resum_expr:gamma_tau} and \eqref{eq:composite_dimensions}. Nevertheless, the most interesting values are $\Delta_{\ph^4}$ and $\Delta_{\ph^6}$. Henriksson got some non-perturbative RG results from \cite{a:frg_Codello_2015}:
        \begin{align}
            \Delta_{\ph^4}|_{d=2.75 ~(\ep=0.125)} &\simeq 1.77; \\
            \Delta_{\ph^4}|_{d=2.5 ~(\ep=0.25)} &\simeq 1.62.
        \end{align}
        Additionally, Henriksson independently calculated the composite operator value in $d=2.75$
        \begin{equation}
            \Delta_{\ph^4}^{\text{Henriksson}}|_{d=2.75} \simeq  1.764.
        \end{equation}
        Using \eqref{resum_expr:gamma_lambda} and \eqref{eq:composite_dimensions} we get:
        \begin{align}
            \Delta_{\ph^4}^{\text{Pade}}|_{d=2.75} &\simeq 1.742; \\
            \Delta_{\ph^4}^{\text{Pade}}|_{d=2.5} &\simeq 1.497.
        \end{align}
        For clarity, we present the same results in Table \ref{tab:phi4_dim}.
        
        \renewcommand*{\arraystretch}{1.4}
        \begin{table}[h!]
        \centering
        \captionsetup{justification=centering}
            \begin{tabular}{c|c|c}
                Methods & $d = 2.75$ & $d = 2.5$ \\
                \hline
                non-perturbative RG \cite{a:frg_Codello_2015} & 1.77 & 1.62 \\
                conformal bootstrap \cite{Henriksson:2025kws} & 1.764 & -- \\
                current work (Pade [2/1]) & 1.742 & 1.497 \\
            \end{tabular}
            \caption{
            Tricritical dimensions of the composite operator $\ph^4$ \eqref{eq:composite_dimensions} in $d=2.75$ ($\ep=0.125$) and $d=2.5$ ($\ep=0.25$) by different methods.
            }
            \label{tab:phi4_dim}
        \end{table}
        One can see that in the $d=2.75$ situation all the results are close to each other, but in $d=2.5$ the match is worse. Unfortunately, data for $\Delta_{\ph^6}$ absent.

\section{Conclusion}\label{sec:conclusion}
   {
    We performed renormalization group calculations in the $\ph^4+\ph^6$ model to third order in the $\ep=(3-d)/2$ expansion. The $\ph^4$ interaction was treated as a composite operator, and its contribution to the renormalization of the $\tau \ph^2$ term was taken into account. The exponent $\omega$, which determines the stability of a fixed point, was calculated as a series in $\ep$. After resummation, $\omega$ is positive in the interval $d \in [2,3)$, indicating IR stability of the fixed point found in this work.
    }

    {
    The expressions for the exponents $\eta$~\eqref{expr:eta} and $\nu$~\eqref{expr:nu} fully coincide with the results of previous works~\cite{vasiliev, LewisAdams, Hager99, Hager02}. The six-loop contribution to the parameter $b_0$, however, differs from that reported in \cite{Hager02}. Computed expressions allow one to calculate the tricritical dimensions of composite operators $\Delta_{\ph^k}$ for $k = 1,2,4,6$. Their exact values are known at $d=2$ from conformal field theory~\cite{BELAVIN1984333}. In addition, some results for intermediate dimensions have been obtained using conformal bootstrap~\cite{a:bootstrap2021, Henriksson:2025kws} and non-perturbative RG~\cite{a:frg_Codello_2012, a:frg_Codello_2015}. Comparison with these studies shows satisfactory agreement in view of the limited number of known terms in the $\ep$ expansion.
    }

    {
    The contribution of the $\ph^4$ interaction to the renormalization of the $\tau \ph^2$ term allowed us to calculate the parameter $a$~\cite{vasiliev}. The sign of $a$ determines which type of behavior is realized along the trajectories with $b<b_0$, where tricritical behavior does not occur. The factor $1/(1-2b_0)$ in $a$~\eqref{a} is highly sensitive to deviations of $b_0$ from its canonical value $1/2$. For small $\ep$, one has $b_0 \simeq 1/2 + 1/5 \ep$, so that $1/(1-2b_0) < 0$. The Pade approximant $\delta \gamma_\tau^*(\ep)$ \eqref{resum_expr:delta_gamma_tau} remains negative throughout the interval $0 < \ep \leq 1/2$. Since $\nu > 0$ in \eqref{a}, it follows that $a > 0$, and modified critical behavior occurs (this situation was discussed in \cite{vasiliev}). On the other hand, the exact value of $b_0$ at $d=2~(\ep=1/2)$ can be obtained from \eqref{eq:nu} -- \eqref{eq:b0}. This yields $b_0|_{d=2} = 4/9$ and therefore $1/(1-2b_0)=9>0$. As a consequence, $a<0$ at $d=2$, and combined tricritical behavior is therefore realized for the trajectories with $b<b_0$. Thus, the quantity $1-2b_0$ vanishes at some intermediate $\ep$ in the interval $0 < \ep < 1/2$, and the behavior changes from modified critical to combined tricritical at this point.
    }
    
    {
    We resummed the parameter $b_0$ in two ways (see \eqref{resum:b01} and \eqref{resum:b02}). The result depends strongly on the choice of resummation. Consequently, higher-order terms of the $\ep$ expansion need to be calculated to achieve more accurate result for $b_0$ (as well as $\delta \gamma_\tau^*(\ep)$ and other exponents).
    }

    {
    In our study, we do not present the results of the diagrams we calculated, which were used to obtain $Z_{1}, \ldots, Z_{4}$. Nevertheless, our results agree with those presented in \cite{vasiliev, Hager99, JackJones}, and some of them can be found in \cite{bednyakov:2025}.
    }

\section*{Acknowledgements}
We thank A. Pikelner for providing analytical results for t-bubble diagrams, A. Bednyakov and M. Nalimov for general discussions. We would also like to thank E. Zerner-Käning for careful reading and editing.  The work is supported by the Ministry of Science and Higher Education of the Russian Federation (agreement no. 075–15–2022–287). M.V.K. gratefully acknowledges the support of the Foundation for the Advancement of Theoretical Physics ``BASIS'' through Grant 25-1-2-48-1.

\subsection*{Conflict of Interest}
The authors declare that they have no conflicts of interest.

\bibliography{references}

\end{document}